\documentclass{ws-ijmpc}

\usepackage{array,booktabs}
\usepackage{bbding}

\begin{document}

\markboth{Peng LUO, Yongli LI, and Chong WU}
{Towards Cost-efficient Sampling Methods}

\catchline{}{}{}{}{}

\title{TOWARDS COST-EFFICIENT SAMPLING METHODS}

\author{Peng LUO}

\address{School of Management, Harbin Institute of Technology, Harbin 150001, P.R.China\\
luopeng\_hit@126.com}

\author{Yongli LI}

\address{1. School of Management, Harbin Institute of Technology, Harbin 150001, P.R.China;\\
2. Dipartimento Di Economia Politica E Statistica, Universit\`{a} Di Siena, Siena 53100, Italy\\
yongli.0440004@gmail.com (corresponding author)}

\author{Chong WU}

\address{School of Management, Harbin Institute of Technology, Harbin 150001, P.R.China; \\
wuchong@hit.edu.cn}

\maketitle

\begin{history}
\received{May 22, 2014}
\end{history}

\begin{abstract}
The sampling method has been paid much attention in the field of complex network in general and statistical physics in particular. This paper presents two new sampling methods based on the perspective that a small part of vertices with high node degree can possess the most structure information of a network. The two proposed sampling methods are efficient in sampling the nodes with high degree. The first new sampling method is improved on the basis of the stratified random sampling method and selects the high degree nodes with higher probability by classifying the nodes according to their degree distribution. The second sampling method improves the existing snowball sampling method so that it enables to sample the targeted nodes selectively in every sampling step. Besides, the two proposed sampling methods not only sample the nodes but also pick the edges directly connected to these nodes. In order to demonstrate the two methods' availability and accuracy, we compare them with the existing sampling methods in three commonly used simulation networks that are scale-free network, random network, small-world network, and two real networks. The experimental results show that the two proposed sampling methods perform much better than the compared existing sampling methods in terms of sampling cost and obtaining the true network structural characteristics.

\end{abstract}

\ccode{PACS Nos.: 02.10.Ox, 02.60.Cb}

\section{Introduction}

Recently, many scholars have investigated a huge amount of studies on complex network in various fields such as computer science$^{1-2}$, statistical physics$^{3-4}$, biology$^{5}$, and sociology$^{6}$. The research on complex network offers a framework for benefiting the structure analysis such as protein interaction network$^{7}$, scientific collaboration networks$^{8}$, and even connections between diseases and symptoms$^{9}$. Thus, capturing the structure characteristics of real-world networks is critical in many research studies, but it would be difficult to obtain the structure characteristics of a network especially when we face the large scale networks such as the World Wide Web whose number of nodes can be millions or even billions$^{10}$. Accordingly, to deal with the above mentioned problem, many sampling methods have been proposed, among which the basic and early ones are Node sampling (NS)$^{11}$ and Edge sampling (ES)$^{12}$. Although recent years have witnessed the ongoing development of network sampling methods, the existing studies suffer from a shortcoming, to our best knowledge, that is to focus on more the accuracy than the efficiency of a sampling method. For example, Ebbes, et al (2012) evaluated the accuracy of nine different sampling algorithms in recovering the underlying structural characteristics of the studied networks$^{13}$. However, the problem of the sampling efficiency is also not further discussed. Here, we consider that a sampling method has a higher efficiency if it needs a lower sampling rate to achieve a given desirable accuracy. In this paper, we focus on proposing two sampling methods which not only own the lower sampling cost but also enjoy higher ability of obtaining the true network structure information.

The proposed methods in this paper would have one distinctive feature compared with these existing classic ones. It is the feature that our complementary approaches are highly efficient in sampling the nodes with high degree values. Although the high degree nodes often occupy a small part of the network nodes, they possess the most structure information of the whole network according to the 80/20 rule$^{14}$. The first proposed sampling method, called improved stratified random sampling (ISRS) method, is based on the existing stratified random sampling (SRS)$^{15}$ method. We use the idea from SRS to classify the nodes based on their degree distribution and then find the targeted nodes with high degree values. Next, we sample the high degree nodes with the higher probability and select the low degree nodes with a lower probability. As a result, ISRS has an improved ability of sampling these nodes which contains more network structure information. We also develop the classic snowball sampling$^{16-17}$ and propose the second new sampling method called improved snowball sampling (ISBS) method. The whole process of ISBS is similar to the snowball sampling, except that it chooses the nodes whose degree is larger than the other half in each step while the original snowball sampling method just picks all the nodes in each step. Besides, we also sample the nodes as well as the edges directly linked to those nodes in the two proposed sampling methods. In all, these two sampling methods can mine and sample the high degree nodes effectively, which allows us to obtain the network structure properties by sampling methods with higher accuracy and lower cost. Here, the low sampling cost means that a low sampling probability can be adopted to obtain much information about the network structure.

In order to demonstrate the method's reasonability and express the work clearly, the rest of this paper is organized as follows. In the following part 2, the related work is reviewed in brief. In part 3, the two new proposed sampling algorithms are explained and showed. In part 4, based on the three commonly used simulation networks, the comparisons between the two proposed methods and some selected existing classic methods are provided to illustrate the new methods' priority. In part 5, the tests in two real networks are provided further. Section 6 concludes and discusses the future work.

\section{Related work}

In our opinion, four kinds of sampling methods can be regards as the cornerstone of the existing numerous network sampling methods. The four basic methods are node sampling (NS)$^{11}$, edge sampling (ES)$^{12}$, snowball sampling (SBS)$^{16-17}$ and random walk sampling (RWS)$^{18}$, and each one bears its own distinctive features. We next introduce them in brief one by one.

\textbf{Node sampling (NS)$^{11}$}. NS method chooses nodes independently and uniformly from the original network. Each node is sampled in such a probability that is a target fraction of nodes required. The sampled network consists of the selected nodes as well as the edges related to the sampled nodes. Besides, many other types of NS methods have been proposed in recent years. A commonly used node sampling approach is the stratified random sampling (SRS)$^{15}$, where nodes are partitioned into different categories and are randomly sampled in different groups$^{19}$.

\textbf{Edge sampling (ES)$^{12}$}. As for ES method, it focuses on edges rather than nodes to consist the sample set. This method chooses the edge randomly and also the two nodes linked by the edge to form the sample set. In recent years, Ahmed et al (2010) developed ES method by introducing graph induction and adopted it to exploits temporal clustering often found in real social networks$^{11}$.

\textbf{Snowball sampling (SBS)$^{16-17}$}. SBS samples nodes by using breadth-first search that starts from a random root node. All nodes linked to the root node are chosen, and then all nodes directly connected to those picked vertices are selected until the desired sample size is achieved. SBS is a widely used sampling method, for example Illenberger et al. (2011) successfully applied SBS to explore the individuals' characteristics and their spatial structure in a real network$^{20}$.

\textbf{Random walk sampling (RWS)$^{18}$}. RWS method takes advantage of the natural connectivity of the network to sample nodes and edges. In RWS, a random seed node is appointed firstly as the hop node in the first step, and then a hop node is picked randomly from the neighbors of the hop node in the last step. The sampling process continues until the desirable number is reached. There are also many developed RWS methods being proposed in recent years. The famous Metropolis-hasting random walk sampling$^{21}$ is one of them which samples nodes by introducing the Metropolis-hasting algorithm to improve the quality of sampling. Similar to SBS, RWS is also a common method in the field of network sampling. For example, Lu and Li (2012) took advantage of RWS to estimate the properties of large online social networks and showed that its results were much better than those based on NS method or ES method$^{22}$.

With time going by, two trends can be summarized on the basis of the above listed four methods: one is to improve the basic ones to create some new sampling methods by combining them or by introducing some new techniques from other disciplines, and the other is to compare these basic methods in terms of different indexes of network statistical properties. We will briefly review some examples as to the two aspects in the following parts.

\textbf{Derivations of the basic methods}. Ribeiro and Towsley (2010) proposed an $m$-dimensional random walk sampling method called frontier sampling (FS)$^{23 }$based on the basic RWS method. Gao et al (2014) introduced the random multiple snowball with cohen process sampling by developing the existing SBS method$^{24}$. Also, Rezvanian et al (2014) utilized the distributed learning automata to sample the complex networks$^{25}$, whose method can be regarded as a method combining the features of NS method and ES method. There are also many similar examples which we have not mentioned. In fact, it is impossible for us to mention all of them one by one. Note that the proposed methods in this paper aims to improve SBS and RWS methods, which can be seen as the derivations of the basic methods, thus this paper follows the first mainstream in the field.

\textbf{Indexes of network statistical properties and Comparisons between the basic methods}. Generally speaking, the commonly used indexes of network statistical properties includes degree distributions, clustering coefficient, closeness centrality, betweenness centrality, Bonacich centrality, the shortest path length, assortativity and so forth. The existing studies often choose some of the above mentioned indexes for comparing the accuracy of the basic methods. For example, Lee et al (2006) investigated the accuracy of Random walk sampling (RWS) and snowball sampling (SBS) in such indexes of network structural characteristics as degree distribution, assortativity, clustering coefficient and betweenness centrality$^{26}$, and Son et al (2012) compared RWS and SBS on statistical properties including degree and assortativity of the directed networks $^{27}$. In this paper, we, following the existing literatures, will choose some of the common indexes to compare the proposed new methods with some selected classic ones. Besides, we will highlight the sampling efficiency in the comparisons.

\section{Method}

As we have mentioned in the Introduction, sampling is a useful approach to analyze the large-scale data set, and a proper sampling method should keep the important and core information that can be extracted from the output data$^{28}$, which is one of critical objectives of a sampling algorithm. On the other hand, generally speaking, when the sampling probability is low, it is challenging for us to find a good sampling method to acquire the correct structure information of the networks. To deal with this problem, we recall the famous 80/20 rule proposed by Pareto$^{29}$, which tells us that a small fraction of nodes can possess the most structure information of the network. Based on this idea, we aim to create two new sampling algorithms which highlight the utilization of some important nodes such as the ones with high in-degrees or out-degrees, the ones with high Katz centrality scores, and so forth. How to mine the important nodes by sampling methods would be the first step for our new method. In this part, we will propose two new sampling methods which not only can keep and uncover the core important information, but also enable to cost a low sampling probability.

Note that our methods are suitable for undirected networks. To make the following expression more concise and clear, we firstly give the network's mathematical expression as follows: Considering an undirected network $G$ with $n$ nodes, it can be described by the symmetric matrix ${\mathbf{A}} = {[{a_{ij}}]_{n \times n}}$, where $${a_{ij}} = \begin{cases}\text{weight between nodes}\;i\;\text{and}\;j,&\quad\text{if there exists a link between the nodes} \\ 0,&\quad\text{otherwise} \end{cases},$$ Especially, as for an unweighted network, $${a_{ij}} = \begin{cases} 1,&\quad\text{if there exists a link between nodes}\;i\;\text{and}\;j \\ 0,&\quad\text{otherwise} \end{cases}.$$

\subsection{Improved stratified random sampling method (ISRS)}

We use the idea of stratified sampling to find the important nodes and accordingly propose our new sampling method. It would be easier for us to sample the important nodes when the nodes are classified into groups based on the Katz centrality, which is the core of our idea. Our method has the premise that the nodes in the group with the higher Katz centrality scores are consider to be more important than those with the smaller scores in terms of uncovering the structure information of a network. To make our study sound, we will demonstrate the idea and the premise in the following numerical experiments and real applications in this paper.

\begin{figure}[!htb]
    \centering
    \includegraphics[scale=1]{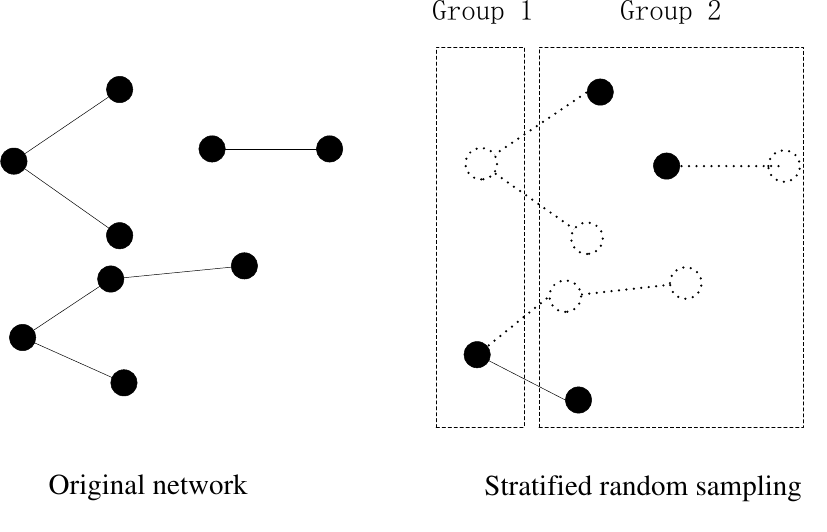}\\
    \caption{\label{fig:1}The basic principle of stratified random sampling}
\end{figure}

Next, we firstly introduce the existing stratified random sampling method which contains the idea of stratified sampling and then discuss how to improve it in aim of enabling the method to cost a low sampling probability given the same accuracy. The process of stratified random sampling is shown in \figurename{} \ref{fig:1}, where is shown that stratified random sampling method starts by classifying the sampling units into a set of mutually exclusive and collectively exhaustive groups$^{15}$. Based on these determined groups, we are able to obtain a simple random sample independently from each stratum and then combine the samples to form a whole stratified random sample. The sampling size from each group is proportionate to the size of corresponding stratum$^{19}$.

To improve the existing stratified random sampling method, we design a series of different sampling probabilities for different stratums which are obtained according to the estimated nodes' degrees. In detail, we first apply a simple random sampling method to sample a little fraction of the network nodes and obtain the node degree of the sampling nodes. Based on the degree distribution of the sampled nodes, we can get the threshold value which equals the top one-fifth quantile according to 80/20 rule. Then, the sampled nodes are divided into two stratums by the obtained threshold value. In the first stratum containing the nodes with higher degrees, we further sample these nodes with a higher probability such as 0.8. Meanwhile, in the second stratum containing the nodes with lower degrees, we further sampled these nodes with a lower probability such as 0.3. Note that we find the above two probability numbers are more proper by numerical experiments through many networks, which can also be demonstrated by the numerical results in the following sections of this paper. Finally, we examine the sampled nodes obtained in this step carefully and achieve all the edges directly linked with these nodes. In fact, it is practical in the real world for surveying some selected nodes carefully.

Given the sampling ratio$x$in the first step, we can find the sampling rate of this method is 0.4$x$, although random sampling in the first step also needs some time and budget. The above sampling rate can be calculated from $0.8 \cdot 0.2 \cdot x + 0.3 \cdot 0.8 \cdot x$, in which the first part means the sampling in the first stratum and the second part means that in the second stratum. In order to present the whole sampling process clearly, \tablename{} \ref{tab:1} shows the corresponding algorithm of the improved stratified random sampling (ISRS, for short) method.

\begin{table}[!htb]
    \centering
    \tbl{Algorithm of ISRS}{
    \begin{tabular}{p{\linewidth}}
        \toprule
        \textbf{\underline{Input}}: the given undirected network ${\mathbf{G}}$ (\textit{n nodes and its matrix expression ${\mathbf{A}}$);}\\
        \hspace{8.5ex}the sampling ratio $x$;\\
        \textbf{\underline{Output}}: the sampled network ${\mathbf{\tilde G}}$; the sampling size $n'$;\\
        \textbf{\underline{Initialization}}:\begin{itemize}
            \item[(1)] $n' \leftarrow 0$ and $a(n,n) \leftarrow 0$.
            \item[(2)] Randomly sampling the given network ${\mathbf{G}}$with the sampling ratio $x$, and then getting each sampled node's degree denoted by ${d_i}$.
            \item[(3)] Calculating the threshold value $c$ of the sampled node degrees according to 80/20 rules.
        \end{itemize}\\
        \textbf{\underline{Loop}}:\\
        \addtolength\leftskip{2em}for  $i$ =1 to $n$\par
        \quad if ${d_i} \geqslant c$ and $rand(1) \leqslant 0.8$\par
        \qquad then $a(i,i) \leftarrow 1$; $n' \leftarrow n' + 1$;\par
        \quad elseif ${d_i} < c\;\& \;rand(1) \leqslant 0.3$\par
        \qquad then $a(i,i) \leftarrow 1$; $n' \leftarrow n' + 1$;\par
        \quad else\par
        \qquad CONTINUE;\par
        \quad end\par
        end\\
        \textbf{\underline{print}}: the sampled network ${\mathbf{G'}} = a \cdot {\mathbf{A}}$ and the sampling size $n'$.\\
        \bottomrule
    \end{tabular} \label{tab:1}
    }
\end{table}

\subsection{Improved snowball sampling method (ISBS)}

\begin{figure}[!htb]
    \centering
    \includegraphics[scale=1]{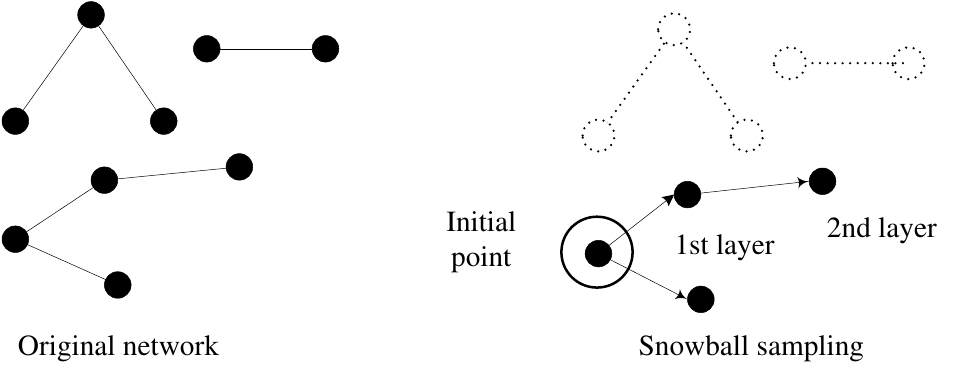}\\
    \caption{\label{fig:2}The basic principle of snowball sampling method}
\end{figure}

As for the existing snowball sampling mathod$^{16-17}$, its process is as follows: a single node is chosen firstly and then we choose all the nodes directly linked to it. Next, all the nodes connected to those selected vertices are picked. The procedure continues until the number of sampling nodes is enough or called meet the budget. The \figurename{} \ref{fig:2} illustrates the process of the snowball sampling, where the set of nodes picked in the $n$th step is denoted as the $n$th layer. As the above process shows, the snowball sampling method tends to choose hubs (nodes with many links) due to high connectivity of them$^{30}$.

However, given a very low sampling probability, it would be not enough for the existing snowball sampling method to choose the nodes with many links. Thus we present the improved snowball sampling (ISBS) method which also emphasizes the nodes with high degrees. In each step, we choose half of the nodes in this layer whose degree is larger than the other half. To control the number of the sampling nodes, we choose the nodes according to the order of their degrees in the terminal layer. Besides, these sampled nodes are also given carefully examined by achieve all the edges directly linked with them, which is similar to the last step of ISRS. The whole process of our improved method ISBS is the same with snowball sampling method expect the part of choosing the nodes with high degrees in each layer, so we just highlight this part in the following algorithm shown in \tablename{} \ref{tab:2}.

\begin{table}[!ht]
    \centering
    \tbl{Algorithm of the highlighted sampled part}{
    \begin{tabular}{p{\linewidth}}
        \toprule
        \textbf{\underline{Input}}: The sampled network's matrix expression ${\mathbf{A}}'$ in this layer;\\
        \hspace{8.5ex}the set of the nodes in the $n$th layer ${b_n}$;\\
        \hspace{8.5ex}the number of the nodes in in the $n$th layer ${t_n}$;\\
        \textbf{\underline{Output}}: the set of sampling nodes in the $n$th layer ${b_n}^\prime $;\\
        \textbf{\underline{Initialization}}:\\
        \addtolength\leftskip{2em}(1) $d\left( {1,{t_n}} \right) \leftarrow 0$;\\
        \textbf{\underline{Loop}}:\\
        \addtolength\leftskip{2em}for $i = 1$ to ${t_n}$\par
        \quad $d\left( {1,i} \right) = sum({\mathbf{A}}'(,:))$;\par
        end\par
        $ma \leftarrow \max (d)$, $mi \leftarrow \min (d)$, $m \leftarrow \left( {ma - mi} \right)$;\par
        for  $i = 1$ to $m$\par
        \quad $mi \leftarrow mi + 1$; $k \leftarrow 0$;\par
        \quad for $j = 1$ to ${t_n}$\par
        \qquad if ${d_j} \leqslant mi$\par
        \qquad\quad then  $k \leftarrow k + 1$;\par
        \qquad end\par
        \qquad if $k/{t_n} \geqslant 0.5$\par
        \qquad\quad then break;\par
        \qquad end\par
        \quad end\par
        end\par
        $j \leftarrow 0$;\par
        for $i = 1$ to ${t_n}$\par
        \quad if ${d_i} \geqslant mi$\par
        \qquad $j \leftarrow j + 1$;${b_j}^\prime  \leftarrow {b_i}$;\par
        \quad end\par
        end\\
        \textbf{\underline{print}}: the sampled network ${b_n}^\prime $ and the sampling size in the $n$th layer $j$.\\
        \bottomrule
    \end{tabular}\label{tab:2}
    }
\end{table}

\section{Numerical Experiments}

To verify these proposed sampling methods, we focus on comparing them with the other classic sampling methods in three commonly used simulation networks, i.e., random network, small-world network and scale-free network. Random network was proposed by Erd\"{o}s and Reyi (1959)$^{31}$ who presented a way to generate network by connecting every pair of nodes with independent probability $p$. As for small-world network, Watts and Strogatz (1998)$^{32}$ constructed it by rewiring each edge randomly with probability $p$ from a regular network with $N$ nodes and $k$ edges. Because small-world network owns some special structure properties, i.e., small diameter and high degree of clustering, it is more similar to real network than random network. Unlike random network and small-world network that lack a hub characteristic of well-connected nodes, scale-free network is prior in terms of exhibiting the occurrence of highly connected nodes. Note that many real networks own the highly connected nodes, such as the collaboration network, the social network and so on. Scale-free network has been studied by Barab\'{a}si and Albert (1999)$^{33}$ and can be generated by adding edges to the given nodes in order to meet a power-law distribution.

Several widely used indexes of network statistical properties have been introduced in related work. Here, we choose three from them for testing and comparing the mentioned sampling methods. The selected three indexes are clustering coefficient, Bonacich centrality and average path distance. Why do we choose them together rather than the others? Firstly, Clustering is associated with a network's local effect$^{18}$. Clustering coefficient enables to measure the strength of linking between a node and its nearest neighbors in a network. To demonstrate that our new sampling methods' ability of reflecting a network's local information, we select clustering coefficient as an index to compare our new methods with the other sampling methods. Besides, we also select another measure to reflect the influence of a node by considering its position and neighbors, in other words the medium-level information of a network. Bonacich centrality$^{34 }$is such an index that indicates a node's influence by considering its connecting nodes and its own position$^{35}$. Furthermore, we also choose average path distance$^{36}$ to measure the global information of a network. In all, the three indexes are selected to test the mentioned sampling methods from micro, medium, and macro perspective so that the test could be a sound one and would provide a confident conclusion.

Based on the three kinds of simulation networks, the sampling rates are designed to change from 0.12 to 0.20 by 0.02 in each step. In each step, we repeat the sampling 500 times. Also, all the simulation networks have 500 nodes. In the following three parts, we will illustrate the results one by one and explain the listed methods' merits and demerits.

\subsection{Clustering coefficient}

The node $i$'s clustering coefficient ${C_i}$ is defined as the ratio of connected pairs among the pairs of $i$'s nearest neigbors$^{37}$. Its mathematical expression is
\begin{equation} {C_i} = \frac{{2{t_i}}}{{{k_i}\left( {{k_i} - 1} \right)}} \label{eq:1}\end{equation}

where ${k_i}$ is the degree of node $i$ and ${t_i}$ denotes the number of links in its neighbors. Furthermore, we use \textbf{Relative Error (RE)$^{38-39}$} to assess the accuracy of the above mentioned clustering coefficient in the sampled networks. Its definition for node $i$ is
\begin{equation} R{E_i} = \frac{{\left| {{C_i} - {C_i}^\prime } \right|}}{{{C_i}}} \label{eq:2}\end{equation}

where ${C_i}^\prime $ denotes the node $i$'s clustering coefficient in the sampled network and ${C_i}$ is the node $i$'s clustering coefficient in the original network. By averaging all the nodes' $R{E_i}$ values, the whole RE value of a sampled network can be obtained. From the above definition, a good sampling method should have a small RE value.

We compare the whole RE values of different sampling methods in three kinds of simulation networks. Here, we repeat the numerical experiments 500 times at each sampling rate and average the obtained RE values at the same sampling rate. Then, the following figures show the resulting RE values from eight different sampling methods whose short names and the corresponding full names has been mentioned and could be found in related work and the titles of section 3.1 and 3.2.

\begin{figure}[!htb]
    \centering
    \includegraphics[width=\linewidth]{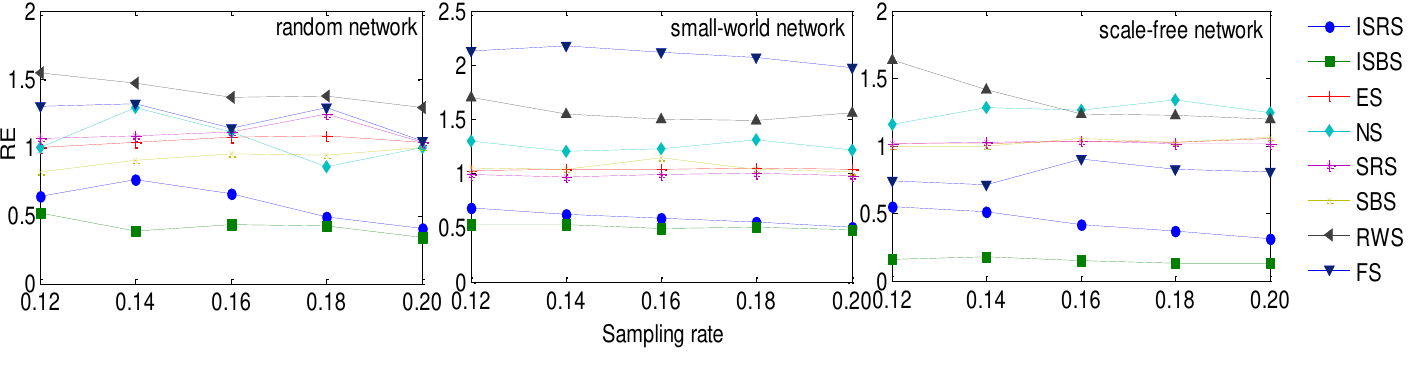}\\
    \caption{\label{fig:3}Results on clustering coefficient}
\end{figure}

As shown in the above linear graphs, the ISBS method displays the overwhelmingly best performance, which is followed by the ISRS method. Furthermore, the sampling rate is low here and as a result, to increase the sampling rate can not greatly uplift the performances of these methods. Even if the sampling rate is low, the performances of the two proposed methods in this paper are not bad, which demonstrates that the two methods are cost-efficient especially when considering the low sampling rates. The conclusions are consistent hereafter because all the experiments here adopt low sampling rates, which is different with the existing literatures because the sampling rates in many existing studies change from 0.2 to 1. Moreover, note that clustering coefficient reflects the local information of a network, thus the two new sampling methods are capable of acquiring the local information of a network.

\subsection{Bonacich centrality}

Bonacich centrality is defined based on the idea that the centrality of a node is affected by its influential neighbors. It not only takes the centrality of the neighboring vertices into consideration, but also considers the position of a node in a network. Given a network with its adjacency matrix ${\mathbf{A}}$, if ${\mathbf{M}} = {\left[ {{\mathbf{I}} - \alpha {\mathbf{A}}} \right]^{ - 1}}$ is well defined and nonnegative, then Bonacich centrality vector of all the nodes is defined as$^{40}$
\begin{equation} {\left[ {{\mathbf{I}} - \alpha {\mathbf{A}}} \right]^{ - 1}} \cdot {\mathbf{1}} \label{eq:3}\end{equation}

where ${\mathbf{I}}$ is identity matrix, $\alpha $ is a scalar and ${\mathbf{1}}$ is a column vector with all its elements as 1. To our best knowledge, \textbf{Pearson's Correlation Coefficient (PCC)$^{41}$} is a good index to measure the similarity between the sampled nodes' Bonacich centrality values and the real values of these nodes in the original network. The performance of a sampling method is much better, if its PCC is much closer to 1. The way of calculating PCC is
\begin{equation} PCC = \frac{{n'\sum\limits_{i = 1}^{n'} {{x_i}{y_i}}  - \left( {\sum\limits_{i = 1}^{n'} {{x_i}} } \right)\left( {\sum\limits_{i = 1}^{n'} {{y_i}} } \right)}}{{\sqrt {n'\sum\limits_{i = 1}^{n'} {x_i^2}  - {{\left( {\sum\limits_{i = 1}^{n'} {{x_i}} } \right)}^2}} \sqrt {n'\sum\limits_{i = 1}^{n'} {y_i^2}  - {{\left( {\sum\limits_{i = 1}^{n'} {{y_i}} } \right)}^2}} }} \label{eq:4}\end{equation}

where ${x_i}$ denotes node $i$'s Bonacich centrality values in the sampled network, ${y_i}$ is node $i$'s Bonacich centrality values in the original network and $n'$ is the node number in the sampled network.

\begin{figure}[!htb]
    \centering
    \includegraphics[width=\linewidth]{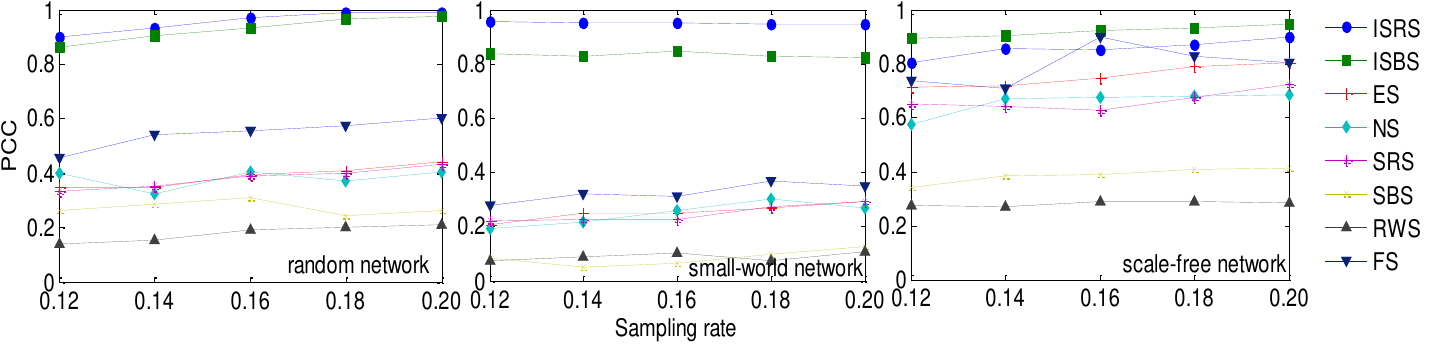}\\
    \caption{\label{fig:4}Results on Bonacich centrality}
\end{figure}

Similar to the process show in section 4.1, the results by averaging 500 experiments are shown in \figurename{} \ref{fig:4}. Among the three kinds of simulation networks, the ISBS and ISRS methods consistently outperform the other approaches. Because our proposed methods are much more efficient to mine the high degree nodes with low sampling rates, our methods are relatively better than other sampling methods in our opinion. Additionally, these sampling methods perform differently in different kinds of networks. For example, the result of ISRS is a little better than ISBS in small-world network, but the opposite phenomenon appears in scale-free network and random network. The reason may be that the ISBS is more skilled at mining the nodes with high degrees which are the feature of scale-free network and are critical for calculating Bonacich centrality. Except scale-free network, the ISRS performs better than the ISBS in obtaining the more accurate Bonacich centrality which reflects the medium-level information of a network.

\subsection{Average path distance}

Average path distance is the average of shortest paths between all the node pairs in a network. The shortest path linking two vertices is their geodesic. If two nodes are unreachable, the shortest path distance between them is infinite and this case is not considered when average path distance is calculated. More information about calculating average path distance can be found in Fronczak et al (2004).

\begin{figure}[!htb]
    \centering
    \includegraphics[width=\linewidth]{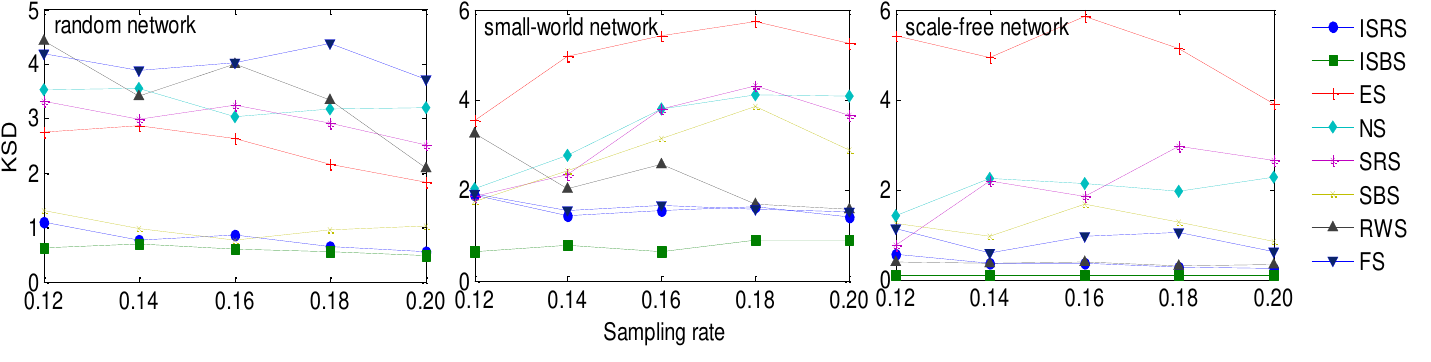}\\
    \caption{\label{fig:5}Results on average path distance}
\end{figure}

\textbf{Kolmogorov-Smirnov D-Statistic (KSD)} is commonly used for calculating the distance between two vectors$^{29}$. Accordingly, this paper adopts KSD to measure the closeness between average path distance in the sampled networks and that in the complete one. Of course, the more close to 0 the calculated KSD is, the better the corresponding method is. Specifically, KSD in this paper can be obtained by the following formula
\begin{equation} KSD = \max \left| {\theta  - \theta _i^s} \right|{  \text{}}(i = 1,2, \cdots ,500) \label{eq:5}\end{equation}

where $\theta $ denotes the average path distance of the original network and $\theta _i^s$ is the obtained average path distance of the sampled network in the \textit{i-}th experiment. Similar to the experiment shown in section 4.1 and 4.2, based on 500 experiments, the average of KSD values at each sampling rate is shown in \figurename{} \ref{fig:5}.

The results show that the superiority of the ISBS in three kinds of simulation networks, although the extent of superiority is different among these networks. Besides, the other methods perform differently in different kinds of networks, but the ISBS is always the best one. The reason for this may be that the ISBS has higher ability of sampling the high degree nodes through which many shortest paths pass in the original networks, while the other method's lower ability may lead to longer and more circuitous routes between nodes in the sampled networks. To sum up, the ISBS enables to uncover the global information of sampled network since average path length is a global index of the network structure characteristic.

In all, the two proposed sampling methods, ISBS and ISRS, performance better than the other selected ones in three kinds of simulation networks. Specifically, as for Bonacich centrality, a medium-level index of network structure characteristic, ISBS is better than ISRS except in scale-free network; While, as for clustering coefficient and average path length, ISRS is prior to ISBS in all three kinds of simulation networks. Besides, one thing should be emphasized here that ISBS has a higher time complexity than ISRS, which can be found from the two listed algorithm processes, but ISRS needs more network information than ISBS because ISRS needs to obtain a threshold probability in its first step. Thus, the two sampling methods that we have presented in this paper have different features so that they could be proper for different cases.

\section{Real Data}

We further compare these sampling methods in a two famous real networks that are the protein interaction network$^{7}$ and the roget network$^{43}$. The protein interaction network is comprised of 2361 vertices representing the proteins and 7128 edges donating the connections of proteins, and the roget network, which is a representation of word connectivity, consists of 1022 vertices and 5075 edges, where a thesaurus is seen as the nodes and their relationships as the links. Note that the scale of the two real networks is much larger than the above analyzed simulation networks. A series of smaller sampling rates (0.02-0.1) are adopted here, which is different from them adopted in the simulation networks. Apart from this, the process of testing and comparing these methods is the same with that in the above numerical experiments.

\begin{figure}[!htb]
    \centering
    \includegraphics[width=\linewidth]{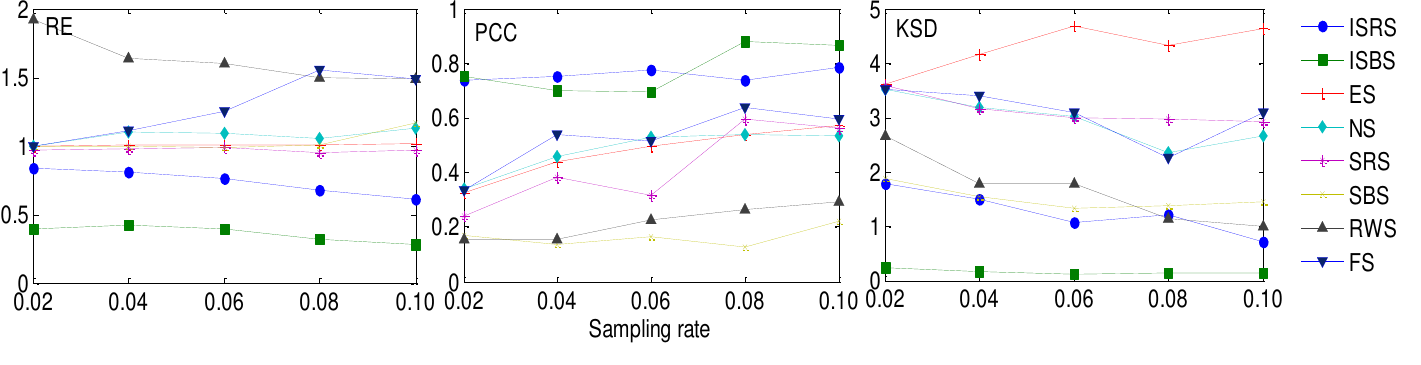}\\
    \caption{\label{fig:6}Results in protein interaction network}
\end{figure}

\begin{figure}[!htb]
    \centering
    \includegraphics[width=\linewidth]{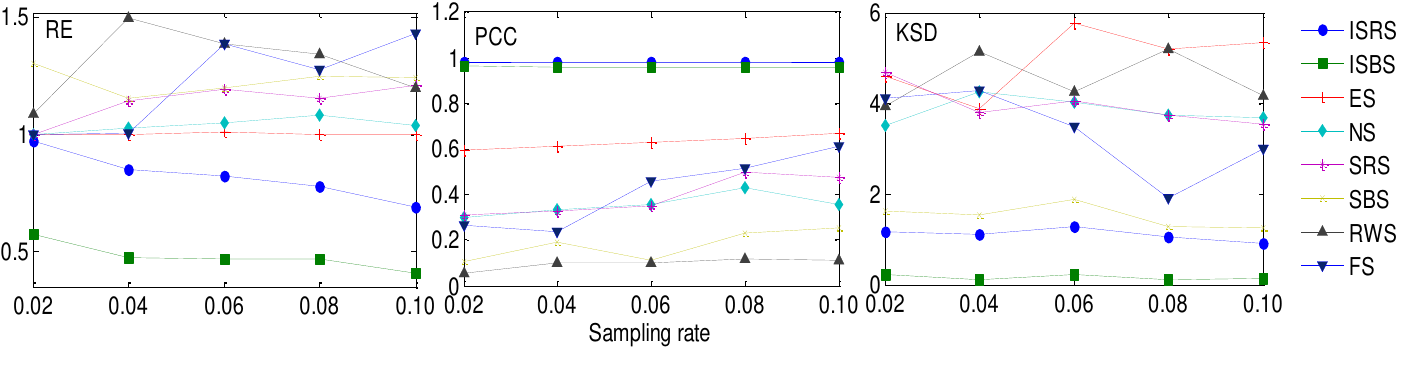}\\
    \caption{\label{fig:7}Results in roget network}
\end{figure}

The results in the two real networks are quite consistent with those in the simulation networks. Here, some smaller sampling rates are adopted and they highlight that the two proposed methods are the cost-efficient ones since they perform not bad even under the small sampling rates. The different performances of two new methods in different indexes illustrate that they can be used for different situations. The tests in real networks validate the applicability of the two new sampling methods even if the scale of the real network is large. What's more important is that the two new methods can uncover the network's structure characteristics in all levels with a low sampling rate. Note that the low sampling rate mentioned here means a low cost of surveying in the real world. Thus, the results indicate that the two new sampling methods are promising ones in the potential numerous applications.

\section{Conclusions and future work}

Our study is based on the perspective that a small part of vertices with high node degree can possess the most structure information of a network. Following this idea, we present two sampling methods which are efficient in sampling the high degree nodes. The first sampling method enhances the classic stratified random sampling and finds the high degree nodes by dividing the nodes based on their degree distribution. Then, we select the high degree nodes with higher probability. By improving the existing snowball sampling method, the second sampling method succeeds in selecting the targeted nodes in each sampling step. Besides, we also acquire the edges directly connected to these sampling nodes in the two sampling algorithms.

We further compare our methods with the existing famous ones such as node sampling, edge sampling, stratified random sampling, snowball sampling, random walk sampling and the recently proposed method, namely frontier sampling. The three network statistical indexes, namely clustering coefficient, Bonacich centrality and average path distance are chosen from the micro, medium, and macro perspective, respectively, in order to evaluate the performance of the methods more soundly. Based on the numerical experiments on three commonly used simulation networks that are scale-free network, random network, small-world network, and two real networks, we can conclude that our methods outperform other sampling methods to some extent. Furthermore, unlike many other studies$^{13,28}$ whose sampling rate are in a much larger scope, our sampling rate is lower than 0.2, which means much lower sampling cost. Furthermore, the two proposed sampling methods have different features: the complexity of ISBS is much higher than that of ISRS, while ISRS should acquire much more information of the network in the sampling process. Thus, they are proper for different cases.

The new proposed methods can be extended in many directions, which could be the future work. Firstly, we could explore the statistical characteristics of our methods' samples and try to correct their biases. Secondly, the method can be applied to find the important nodes. For example, in the coauthor network, the most influential authors can be found by our methods so that we can make better decision to select a collaborator. Thirdly, the communities detecting methodology can be utilized combained with our sampling technology so as to study the true structure characteristics of the communities in a low cost. Last but not the least, as the scale of the real networks becomes larger and larger, the demand of efficient and lower cost sampling technology is stronger and stronger. We deeply hope the method presented here will contribute to it.

\section*{Acknowledgements}

This study was partly funded by National Natural Science Foundation of China (No.71271070) and China Scholarship Council (No.201306120159).


\section*{Supplementary Materials}
From http://www.escience.cn/system/file?fileId=66575, the Main Codes could be downloaded.

\end{document}